\documentclass[12pt]{iopart}
\usepackage{graphicx}
\usepackage{iopams}  
\begin{document}

\title{Energy and temperature fluctuations in the single electron box}

\author{Tineke L. van den Berg, Fredrik Brange, Peter Samuelsson}

\address{Department of Physics, Lund University, Box 118, S-221 00 Lund, Sweden}
\ead{samuel@teorfys.lu.se}
\vspace{10pt} 
\begin{indented}
\item[]February 2015
\end{indented}

\begin{abstract}
In mesoscopic and nanoscale systems at low temperatures, charge carriers are typically not in thermal equilibrium with the surrounding lattice. The resulting, non-equilibrium dynamics of electrons has only begun to be explored. Experimentally the time-dependence of the electron temperature (deviating from the lattice temperature) has been investigated in small metallic islands. Motivated by these experiments we investigate theoretically the electronic energy and temperature fluctuations in a metallic island in the Coulomb blockade regime, tunnel coupled to an electronic reservoir, i.e. a single electron box. We show that electronic quantum tunnelling between the island and the reservoir, in the absence of any net charge or energy transport, induces fluctuations of the island electron temperature. The full distribution of the energy transfer as well as the island temperature is derived within the framework of full counting statistics. In particular, the low-frequency temperature fluctuations are analysed, fully accounting for charging effects and non-zero reservoir temperature. The experimental requirements for measuring the predicted temperature fluctuations are discussed.
\end{abstract}

\vspace{2pc}

\section{Introduction}
A macroscopically large system, thermally coupled to a heat reservoir, is in thermal equilibrium with the reservoir when the system temperature is equal to the one of the reservoir. Microscopic fluctuations of heat between the system and the reservoir will not modify the thermal equilibrium. In smaller, meso or microscopic systems this is not necessarily true \cite{Landau}; individual events of energy transfer between the reservoir and the system can induce fluctuations in time of the system temperature \cite{Chui}. Consequently, microscopic heat fluctuations (no external force applied) can drive the small system away from thermal equilibrium with the reservoir. 
The great interest in nanoscale systems during the last decades has motivated
extensive efforts \cite{Jarzynski,Crooks,Giazotto,Rowe,Hanggi,Seifert} to reformulate or invent novel concepts in order to properly describe the thermodynamics of small systems. Here, as an important example, the concept of temperature fluctuations has led to considerable debate \cite{Kittel,Mandelbrot}.

In mesoscopic or nanoscale solid state systems coupled to electronic reservoirs and kept at low temperatures, quantum tunnelling of electrons between the system and the reservoirs is an important source of individual, random events of energy transfer \cite{Averin,Averin2,Kung,Saira,Koski}. A particularly interesting situation occurs when the electrons in the system interact much stronger with each other than with the lattice phonons, often the case in both metals \cite{Giazotto} and semiconductors \cite{Gaspref}. The electron-electron interactions lead to a rapid thermalization of the electron gas, on a time scale $\tau_{\mathrm{e-e}}$. As a consequence, the electrons are in local thermal equilibrium, characterized by a Fermi distribution with temperature $T_\mathrm{e}$. However, the weak electron-phonon interaction leads to that the electron gas is thermalized with the lattice phonons on a much slower time scale $\tau_{\mathrm{e-ph}}$. If the energy exchange between the system and the reservoir, due to tunnelling, occurs on an intermediate time scale $\tau_{\mathrm{E}}$, i.e. 
\begin{equation}
\tau_{\mathrm{e-e}} \ll \tau_{\mathrm{E}} \ll \tau_{\mathrm{e-ph}}
\label{timescales}
\end{equation} 
the system electron temperature will develop fluctuations, $T_{\mathrm{e}}=T_{\mathrm{e}}(t)$, with the dynamics driven only by quantum tunnelling. 

Electronic temperature fluctuations in mesoscopic systems are typically challenging to measure. However, in a number of experiments, focusing on temperature based photon detection, \cite{Martinis,Cleland,Lehnert,Gasparinetti,Viisanen} the electron temperature $T_\mathrm{e}(t)$ of a small metallic island was monitored in real time following a pulsed excitation. In the very recent experiments in Refs. \cite{Gasparinetti,Viisanen}, temperature fluctuations of the order of $10$~mK away from the lattice temperature $\sim 100$~mK could be detected, approaching the limit of single microwave photon detection. Moreover, the observed relaxation time $\tau_{\mathrm{e-ph}}$ of $T_{\mathrm{e}}(t)$ towards the lattice temperature was of the order of $100~\mu$s, several orders of magnitude longer than the typical $\tau_{\mathrm{e-e}}$, of the order of $1$~ns or below \cite{Pothier}. This large separation of time scales puts in prospect experiments with real time monitoring of $T_{\mathrm{e}}(t)$, driven by electron tunnelling occurring on an intermediate time scale $\tau_{\mathrm{E}}$, fulfilling the inequality in Eq. (\ref{timescales}). 

In this work we present a theoretical investigation of the tunnelling-induced electronic heat transfer and temperature fluctuation statistics in a conceptually simple system. Motivated by the experiments \cite{Martinis,Cleland,Lehnert,Gasparinetti,Viisanen} we consider a metallic island in the Coulomb blockade regime, tunnel coupled to a reservoir, i.e. a single electron box \cite{box}, see Fig. \ref{fig_system}. We point out that several theoretical investigations of temperature fluctuation distributions have been performed in related systems, e.g. in open, non-interacting islands \cite{Heikilla}, overheated single electron transistors \cite{Laakso}, time driven systems \cite{Altland}, superconducting heterostructures \cite{Laakso2}, and an island with injection of electronic wave packets \cite{Battista}.  In our work we focus on the case with no external force applied to the system, such as a static or time-varying voltage or thermal bias. In fact, due to Coulomb blockade, only zero or one excess electrons are allowed on the island and hence, there can be no net charge transfer between the system and the reservoir. These simplifying assumptions allow us to provide a detailed, analytical description of the statistical properties of the heat transfer and temperature fluctuations.

To provide a physically compelling picture, we first investigate the statistics of energy transfer in thermodynamic equilibrium, when the system electron temperature $T_{\mathrm{e}}$ is constant and equal to the reservoir temperature $T_{\mathrm{L}}$. We show how the fluctuations of transferred energy depend both on the fluctuations in number of tunnelling events during the measurements as well as on the fluctuations in energy transferred in each tunnelling event. Based on the understanding of the energy fluctuations in thermal equilibrium, we analyse the fluctuations of the system temperature away from equilibrium ($T_{\mathrm{e}}\neq T_{\mathrm{L}})$, induced by the fluctuations of the energy transfer. Employing a Boltzmann-Langevin approach \cite{Buttrev,Heikkilabook} for the fluctuation correlations as well as a semiclassical stochastic path integral technique \cite{pilgrambutt,pilgram} for the full distribution of fluctuations, where the quantum tunnelling of electrons act as the generator of fluctuations, we analyse the statical properties of the temperature fluctuations $\Delta T_{\mathrm{e}}(t)=T_{\mathrm{e}}(t)-\langle T_{\mathrm{e}}(t) \rangle$. 

For the magnitude of the fluctuation correlations, of direct experimental relevance, the result is of the canonical form for fluctuations around thermodynamical equilibrium \cite{Landau}
\begin{equation}
\langle \Delta T_{\mathrm{e}}(t) \Delta T_{\mathrm{e}}(t+t')\rangle=\frac{k_{\mathrm{B}}T_{\mathrm{L}}^2}{C_{\mathrm{e}}}e^{-|t'|/\tau_{\mathrm{C}}}, 
\label{Ttcorr}
\end{equation}
where the correlation time $\tau_{\mathrm{C}}=C_{\mathrm{e}}/\kappa$ with $C_{\mathrm{e}}$ the heat capacity of the island and $\kappa$ the thermal conductance of the island-reservoir contact, both taken at the reservoir temperature $T_{\mathrm{L}}$. We note that the instantaneous fluctuations are given by \cite{Landau} $\langle \Delta T_{\mathrm{e}}^2(t)\rangle=k_{\mathrm{B}}T_{\mathrm{L}}^2/C_{\mathrm{e}}$, independent of both the island-reservoir contact properties as well as charging effects. In contrast, the low-frequency correlator $\int dt' \langle \Delta T_{\mathrm{e}}(t) \Delta T_{\mathrm{e}}(t+t')\rangle=2k_{\mathrm{B}}T_{\mathrm{L}}^2/\kappa$, in accordance with the fluctuation-dissipation theorem for heat transport \cite{AvPekola,Heikkilabook}, depend on both contact properties and charging effects via $\kappa$.   

Higher-order temperature correlators as well as the full counting statistics, clearly demonstrating the non-Gaussian nature of the fluctuations, are investigated with the focus on the low-frequency regime. As a general result we find that the temperature fluctuations increase for increasing charging effects, a consequence of both a wider range of energies of tunnelling electrons participating in the heat transfer and larger fluctuations in the number of tunnel events during the measurement.

The rest of the paper is organized as follows. In Sec. 2 the system and model are presented. Then, in Sec. 3 the probability distribution of the energy fluctuations are derived. In Sec. 4 the energy transfer statistics for equal temperature of the system and the reservoir is analysed. Thereafter, in Sec. 5, the distribution of the temperature fluctuations is derived and analysed in detail. Finally, in Sec. 6 we conclude and discuss prospects for an experimental realization of our proposal.

\section{System and model}
\begin{figure}[h]
\flushright
\includegraphics[width=0.9\linewidth]{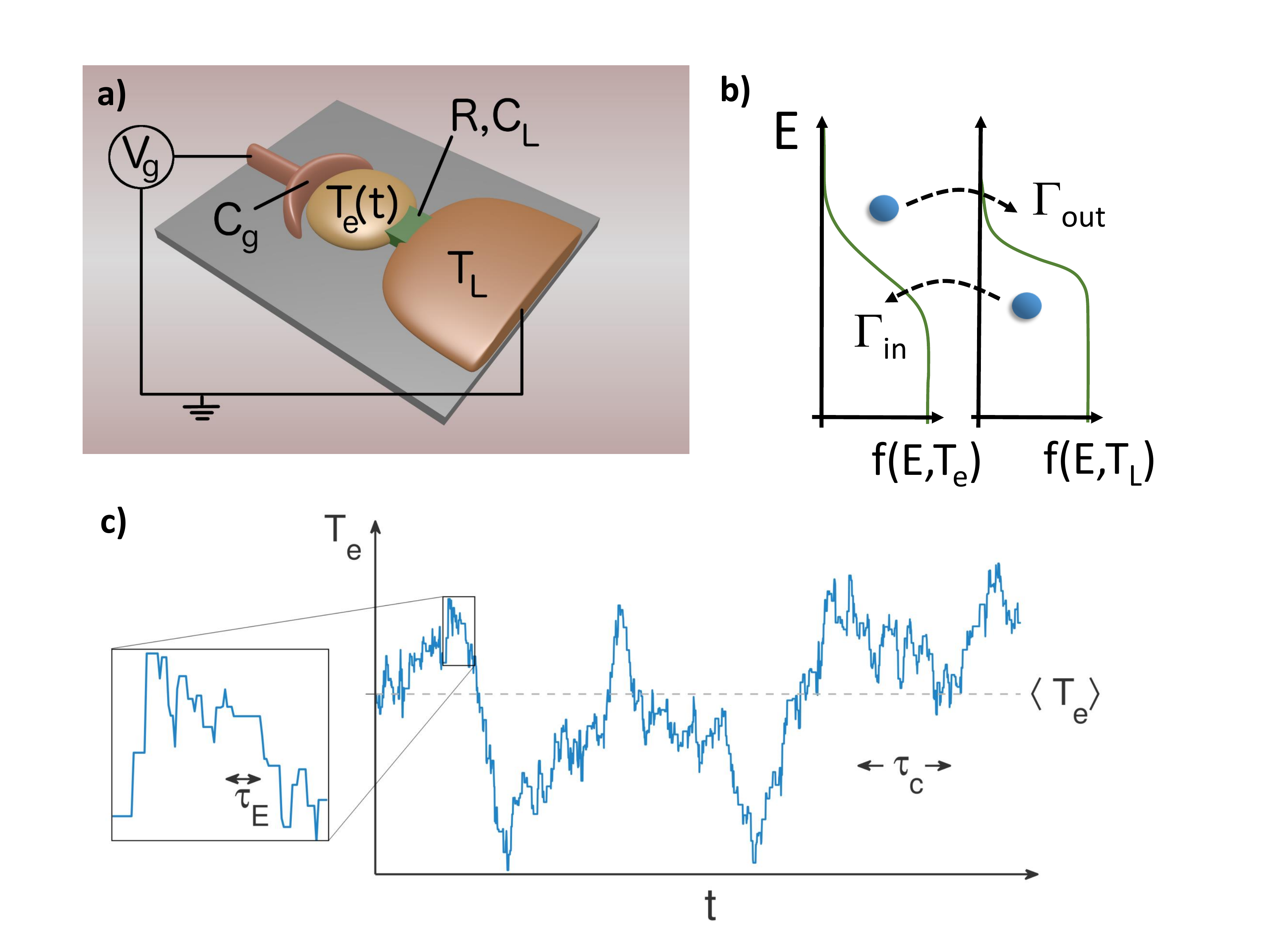}
\caption{a) Schematic of the single electron box. A metallic dot is coupled capacitively, $C_{\mathrm{g}}$, to an electrostatic gate, kept at a voltage $V_{\mathrm{g}}$. The dot is further coupled via a tunnel barrier, resistance $R$ and capacitance $C_{\mathrm{L}}$, to a metallic reservoir. The reservoir, in thermodynamic equilibrium, is grounded and kept at the lattice temperature $T_{\mathrm{L}}$. In the dot, the electrons are in local thermal equilibrium at a temperature $T_{\mathrm{e}}(t)$ fluctuating in time. b) The Fermi distributions of the electrons in the dot (left) and the lead (right), shown schematically. Tunnelling processes in and out of the dot with respective rates, $\Gamma_{\mathrm{in}}$ and $\Gamma_{\mathrm{out}}$, are shown. c) Representative time trace of fluctuation of electron temperature $T_{\mathrm{e}}(t)$. Magnification of short time interval, with individual tunnelling events visible, shown in rectangle. Typical time scales for fluctuation correlations $\tau_{\mathrm{C}}$ and energy tunnelling $\tau_{\mathrm{E}}$ together with average temperature $\langle T_{\mathrm{e}}\rangle=T_{\mathrm{L}}$ shown.}
\label{fig_system}
\end{figure}

We consider the system shown in Fig. \ref{fig_system}. A metallic island, or dot, is coupled via a tunnel barrier, with resistance $R$ and capacitance $C_{\mathrm{L}}$, to an electrically grounded reservoir. The dot is further coupled capacitively, with strength $C_{\mathrm{g}}$, to an electrostatic gate, kept at a potential $V_{\mathrm{g}}$. The dot is in the Coulomb blockade regime, $R\gg h/e^2$, with a charging energy $E_{\mathrm{c}}=e^2/2(C_{\mathrm{L}}+C_{\mathrm{g}})$. The electronic reservoir is kept at thermodynamic equilibrium, at a constant temperature $T_{\mathrm{L}}$, and characterized by a Fermi distribution $f(E,T_\mathrm{L})=1/(1+\exp[E/k_\mathrm{B}T_\mathrm{L}])$ for the electrons, with energy $E$ counted from the Fermi energy. 

On the dot, the electron-electron interaction time $\tau_{\mathrm{e-e}}$ is much shorter than the electron-phonon interaction time $\tau_{\mathrm{e-ph}}$. On time scales longer than $\tau_{\mathrm{e-e}}$, the electron distribution is thermalized, characterized by a Fermi distribution $f(E,T_{\mathrm{e}})$.  The electron temperature $T_{\mathrm{e}}=T_{\mathrm{e}}(t)$ generally depends on time and is typically different from the constant lattice temperature $T_{\mathrm{L}}$ (same as the reservoir temperature). The total dot-reservoir system is thus in a non-equilibrium state, with the dot electrons in quasi-equilibrium.

Electrons tunnel sequentially in to and out of the dot with rates $\Gamma_{\mathrm{in}}^{(0)}$ and $\Gamma_{\mathrm{out}}^{(0)}$ respectively (co-tunnelling is not considered). We consider the situation with $k_{\mathrm{B}}T_{\mathrm{L}},k_{\mathrm{B}}T_{\mathrm{e}} \ll E_{\mathrm{c}}$, and a gate voltage $V_{\mathrm{g}}$ tuned so that only zero or one excess electron on the dot is energetically allowed. The time scale $\tau_{\mathrm{E}}$ for energy exchange between the dot and the reservoir is given by the typical time for an in-tunnelling and a subsequent out-tunnelling (or vice versa), i.e. 
\begin{equation}  
\tau_{\mathrm{E}}=\frac{1}{\Gamma_{\mathrm{in}}^{(0)}}+\frac{1}{\Gamma_{\mathrm{out}}^{(0)}}. 
\end{equation} 
For $\tau_{\mathrm{E}} \gg \tau_{\mathrm{e-ph}}$ the electron-phonon scattering is much faster than the process of energy transfer via tunnelling and the dot electron temperature is constant and equal to the lattice temperature, $T_{\mathrm{e}}=T_{\mathrm{L}}$. The dot-reservoir system is thus in thermal equilibrium. Below, this regime is only considered in order to provide a qualitative understanding of the dot-reservoir energy transfer statistics. The regime of main interest is instead the opposite one, described by Eq. (\ref{timescales}), where the dynamics of $T_{\mathrm{e}}(t)$ is driven by electron tunnelling. We note that the tunnel rates $\Gamma_{\mathrm{in}}^{(0)}$ and $\Gamma_{\mathrm{out}}^{(0)}$ depend on the difference in charging energy for zero and one electrons, further discussed below.

To provide a physically intuitive picture of the mechanism of the temperature fluctuations we consider the following scenario: starting with the dot-reservoir in thermal equilibrium, $T_{\mathrm{e}}=T_{\mathrm{L}}$, an in-and-out tunnelling event take place. The energies of the in (out) tunnelling electrons, added to (subtracted from) the total dot electron energy, are random quantities with probability distributions determined by the dot and reservoir Fermi functions, accounting for the difference in charging energy between zero and one excess electron on the dot. As a consequence, after the tunnelling events the total dot electron energy has typically changed, with the same probability for an increase as for a decrease (as $T_{\mathrm{e}}=T_{\mathrm{L}}$). Since the dot electron temperature is directly related to the total energy, via the heat capacity $C_{\mathrm{e}}$, and the electron gas is thermalized on the short time scale $\tau_{\mathrm{e-e}}$, the dot temperature $T_{\mathrm{e}}$ changes accordingly. Then, with the next in-and-out tunnelling event, $T_{\mathrm{e}}$ changes further and over time the dot electron temperature develops fluctuations, as shown in Fig. \ref{fig_system}. Importantly, for $T_{\mathrm{e}}>T_{\mathrm{L}}$ the probability for a net energy transfer out of the dot during an in-and-out tunnelling event is larger than the probability for a net in transfer, and vice versa for $T_{\mathrm{e}}<T_{\mathrm{L}}$. Hence, there is a negative feedback mechanism inherent in the model, characterized by the time scale $\tau_{\mathrm{C}}$, providing relaxation of $T_{\mathrm{e}}$ towards $T_{\mathrm{L}}$ and hence preventing large deviations of $T_{\mathrm{e}}$ away from $T_{\mathrm{L}}$. 

\section{Energy transfer statistics}

We consider the dot-reservoir energy transfer statistics a time scale $\tau$  long enough for a large number of tunnel events to occur, $\tau \gg \tau_{\mathrm{E}}$, but short enough for the temperature to be taken constant, $\tau \ll \tau_{\mathrm{C}}$. The energy transfer statistics of the dot can be described within the framework of a master equation \cite{BagNaz,Flindt,Kiess,Esposito}, generalized to include energy counting fields $\xi$ \cite{pilgram,kinpil,Esposito},
\begin{equation}
\frac{d}{dt}\left[\begin{array}{c} P_0(\xi,t) \\ P_1(\xi,t) \end{array}\right]=M(\xi)\left[\begin{array}{c} P_0(\xi,t) \\ P_1(\xi,t) \end{array}\right]
\label{ME}
\end{equation}
where $P_0(\xi,t)$ and $P_1(\xi,t)$ are the generalized probabilities for zero and one excess electron on the dot, respectively (note that $P_0(0,t)+P_1(0,t)=1$) and the rate matrix
\begin{equation}
M(\xi)=\left(\begin{array}{cc} -\Gamma_{\mathrm{in}}^{(0)} & \Gamma_{\mathrm{out}}(\xi) \\ \Gamma_{\mathrm{in}}(\xi) & -\Gamma_{\mathrm{out}}^{(0)} \end{array}\right)
\label{ratemat}
\end{equation}
The counting field dependent rates to tunnel in to or out of the dot are given by the standard single charge tunnelling rates \cite{box} modified to include counting field factors as 
\begin{eqnarray}
\Gamma_{\mathrm{in}}(\xi)&=&\frac{1}{e^2R}\int dE  [1-f(E,T_{\mathrm{e}})]f(E+\Delta,T_{\mathrm{L}})e^{i\xi E} \nonumber \\
\Gamma_{\mathrm{out}}(\xi)&=&\frac{1}{e^2R}\int dE  f(E,T_{\mathrm{e}})[1-f(E+\Delta,T_{\mathrm{L}})] e^{-i\xi E}
\label{rates}
\end{eqnarray}
with $\Gamma_{\mathrm{in/out}}^{(0)}=\Gamma_{\mathrm{in/out}}(\xi=0)$.
The energy $\Delta=E_{\mathrm{c}}(1)-E_{\mathrm{c}}(0)=E_{\mathrm{c}}(1-2n_{\mathrm{g}})$ is the difference in charging energy $E_{\mathrm{c}}(n)=E_{\mathrm{c}}(n-n_{\mathrm{g}})^2$ for one and zero excess electrons on the dot and $en_{\mathrm{g}}=V_{\mathrm{g}}C_{\mathrm{g}}$ the gate induced dot charge, with $C_{\mathrm{g}}$ the gate-dot capacitance. 

We are interested in the full distribution of energy, $P_{\tau}(\varepsilon)$ transferred into ($\varepsilon>0$) or out of ($\varepsilon<0$) the dot during the time $\tau$. The distribution is conveniently expressed in terms of a cumulant generating function (CGF) $F(\xi)$ as
\begin{equation}
P_{\tau}(\varepsilon)=\frac{1}{2\pi}\int d\xi e^{-i\xi \varepsilon+F(\xi)}
\label{probdist}
\end{equation}
The CGF is given by $\tau$ times the eigenvalue of the rate matrix $M(\xi)$ in Eq. (\ref{ratemat}) which goes to zero for $\xi \rightarrow 0$, as
\begin{eqnarray}
-2F(\xi)/\tau=\Gamma_{\mathrm{in}}^{(0)}+\Gamma_{\mathrm{out}}^{(0)}-\sqrt{\left(\Gamma_{\mathrm{in}}^{(0)}-\Gamma_{\mathrm{out}}^{(0)}\right)^2+4\Gamma_{\mathrm{in}}(\xi)\Gamma_{\mathrm{out}}(\xi)}
\label{CGF}
\end{eqnarray}
We note that since the temperature in the dot and the lead are typically different, $T_{\mathrm{L}} \neq T_{\mathrm{e}}(t)$, we have no simple analytical expression for the tunnel rates in Eq. (\ref{rates}) and hence not for $F(\xi)$.

\subsection{Lowest order cumulants} 
\begin{figure}[h]
\flushright
\includegraphics[width=0.85\linewidth]{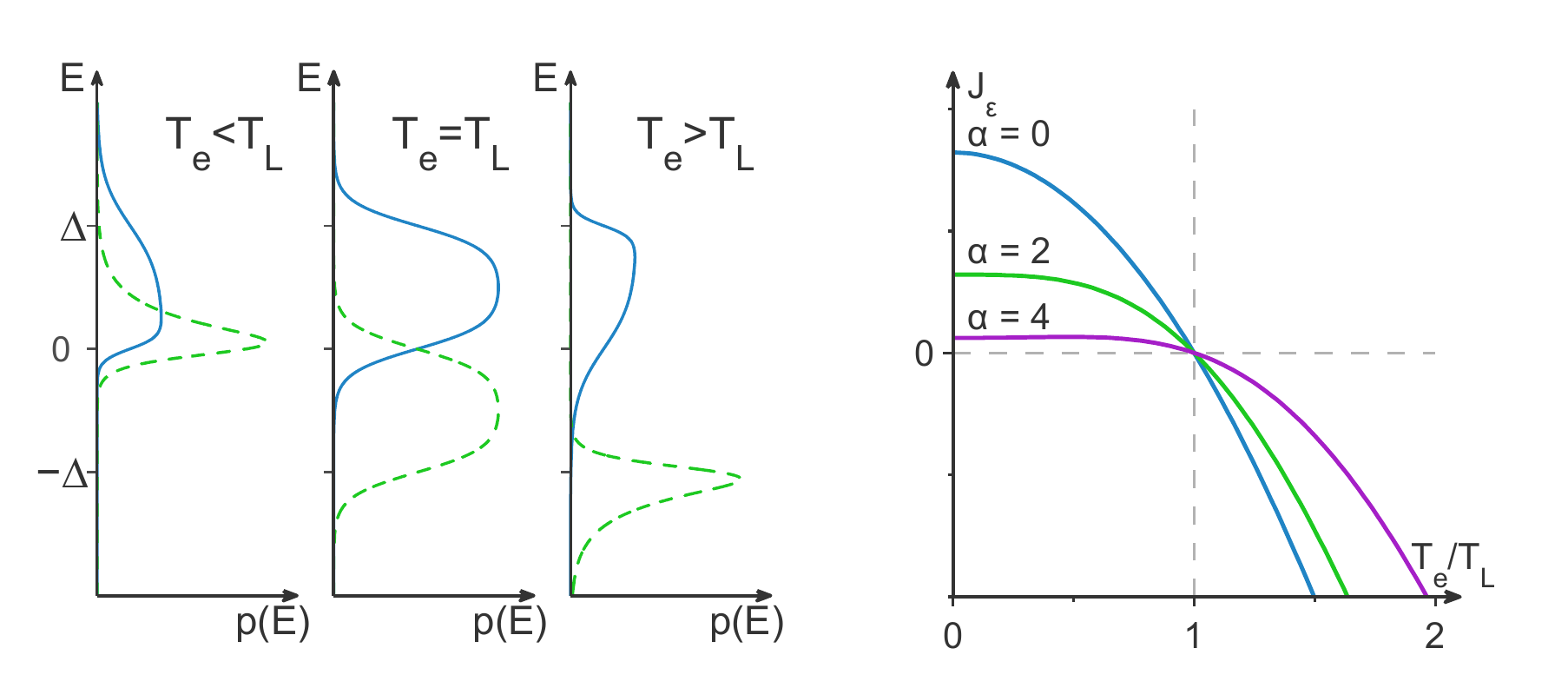}
\caption{Left: Probability distribution $p(E)$ (normalized spectral distribution) for energy of electrons tunnelling between the dot and the reservoir, for $T_{\mathrm{e}}<T_{\mathrm{L}}, T_{\mathrm{e}}=T_{\mathrm{L}}$ and $T_{\mathrm{e}}>T_{\mathrm{L}}$. Solid (dashed) line denotes the distribution for in (out) tunnelling, with $E>0$ for energy added to the dot. Right: Energy current $j_{\mathrm{\varepsilon}}$ (in arbitrary units) as a function of $T_{\mathrm{e}}/T_{\mathrm{L}}$, for $\alpha=\Delta/(2k_{\mathrm{B}}T_{\mathrm{L}})=0,2$ and $4$.}
\label{curr}
\end{figure}

The different cumulants of the energy transfer are given by successive differentiation of the cumulant generating function with respect to $\xi$. The average energy is 
\begin{equation}
\langle \varepsilon \rangle=\left.\frac{dF}{d(i\xi)}\right|_{\xi=0}=\tau\frac{\Gamma_{\mathrm{out}}^{(0)}\Gamma_{\mathrm{in}}^{(1)}-\Gamma_{\mathrm{in}}^{(0)}\Gamma_{\mathrm{out}}^{(1)}}{\Gamma_{\mathrm{out}}^{(0)}+\Gamma_{\mathrm{in}}^{(0)}}
\end{equation}
Here we have introduced for clarity and later convenience
\begin{eqnarray}
\Gamma^{(n)}_{\mathrm{in}}&=&\left.\frac{d^n\Gamma_{\mathrm{in}}}{d(i\xi)^n}\right|_{\xi=0}=\frac{1}{e^2R}\int dE E^n [1-f(E,T_{\mathrm{e}})]f(E+\Delta,T_{\mathrm{L}}) \nonumber \\
\Gamma_{\mathrm{out}}^{(n)}&=&\left.\frac{d^n\Gamma_{\mathrm{out}}}{d(-i\xi)^n}\right|_{\xi=0}=\frac{1}{e^2R}\int dE E^n f(E,T_{\mathrm{e}})[1-f(E+\Delta,T_{\mathrm{L}})]. 
\label{gamman}
\end{eqnarray}
We point out that in the steady state the probabilities obtained from Eq. (\ref{ME}),  for $\xi=0$, are  $P_0^{\mathrm{s}}=1-P_1^{\mathrm{s}}=\Gamma_{\mathrm{in}}^{(0)}/( \Gamma_{\mathrm{out}}^{(0)}+\Gamma_{\mathrm{in}}^{(0)})$ and hence $\langle \varepsilon \rangle=\tau(\Gamma_{\mathrm{in}}^{(1)}P_0^{\mathrm{s}}-\Gamma_{\mathrm{out}}^{(1)}P_1^{\mathrm{s}})$. Since $\Gamma_{\mathrm{in/out}}^{(1)}$ are the energy tunnelling rates, our result is consistent with the definition of average energy current flowing into the dot as
\begin{equation}
j_{\varepsilon}=\frac{\langle \varepsilon \rangle}{\tau}
\label{encurrdef1}
\end{equation}
The current $j_{\varepsilon}$ is plotted as a function of $T_{\mathrm{e}}/T_{\mathrm{L}}$ in Fig. \ref{curr} for a number of different $\Delta/(k_{\mathrm{B}}T_{\mathrm{L}})$, clearly showing that the energy current flows from the hotter to the colder part of the total system and being zero at $T_{\mathrm{e}}=T_{\mathrm{L}}$, as expected. Moreover, the spectral distribution of the energy flow for different temperatures is shown in Fig. \ref{curr}.

The magnitude of the fluctuations of the transferred energy, $\langle \! \langle \varepsilon^2 \rangle \! \rangle=\langle (\Delta \varepsilon)^2 \rangle =\langle (\varepsilon-\langle \varepsilon \rangle)^2\rangle=d^2F/d(i\xi)^2\vert_{\xi=0}$ is given by the expression
\begin{eqnarray}
\frac{\langle (\Delta \varepsilon)^2 \rangle}{\tau}&=&\frac{\Gamma_{\mathrm{in}}^{(2)}\Gamma_{\mathrm{out}}^{(0)}+\Gamma_{\mathrm{in}}^{(0)}\Gamma_{\mathrm{out}}^{(2)}-2\Gamma_{\mathrm{in}}^{(1)}\Gamma_{\mathrm{out}}^{(1)}}{\Gamma_{\mathrm{out}}^{(0)}+\Gamma_{\mathrm{in}}^{(0)}}- \frac{2\left(\Gamma_{\mathrm{out}}^{(0)}\Gamma_{\mathrm{in}}^{(1)}-\Gamma_{\mathrm{in}}^{(0)}\Gamma_{\mathrm{out}}^{(1)}\right)^2}{\left(\Gamma_{\mathrm{out}}^{(0)}+\Gamma_{\mathrm{in}}^{(0)}\right)^3}
\end{eqnarray}
Third and higher order cumulants of the transferred energy are found accordingly. 

\section{Equal temperature case}
To gain a deeper physical understanding of the dot electron temperature fluctuations, it is instructive to first consider the case where the dot electrons equilibrate with the surrounding lattice on a time scale much faster than the average time $\tau_{\mathrm{E}}$ for an in-and-out tunnel event. In this case we have a constant $T_\mathrm{e}=T_\mathrm{L}$, i.e. the reservoir and the dot are in thermal equilibrium during the entire time $t_0$ of the measurement. It is then also possible to obtain an analytical expression for the tunnel rates in Eq. (\ref{rates}) and hence the CGF in Eq. (\ref{CGF}), allowing for a detailed investigation of the energy transport in the system. Making use of the relation $f(E+x)[1-f(E)]=[f(E+x)-f(E)]/(1-e^{x/k_{\mathrm{B}}T})$ and the Fourier transform of the Fermi distribution we find for the rates
\begin{equation}
\Gamma_{\mathrm{in}}(\xi)=\Gamma_{\mathrm{in}}^{(0)}h(\xi), \hspace{0.3 cm} \Gamma_{\mathrm{out}}(\xi)=\Gamma_{\mathrm{out}}^{(0)}h(-\xi),
\end{equation}
with
\begin{equation}
\Gamma_{\mathrm{in}}^{(0)}=\frac{1}{e^2 R}\frac{\Delta}{e^{\Delta/k_{\mathrm{B}}T_{\mathrm{L}}}-1}, %
\hspace{0.5cm} \Gamma_{\mathrm{out}}^{(0)}=\Gamma_{\mathrm{in}}^{(0)}e^{\Delta/k_{\mathrm{B}}T_{\mathrm{L}}},
\end{equation}
and
\begin{equation}
h(\xi)=e^{-i\xi \Delta/2}\frac{2\pi k_{\mathrm{B}}T_{\mathrm{L}} \sin(\xi\Delta/2)}{\Delta \sinh(\pi k_{\mathrm{B}}T_{\mathrm{L}}\xi)}.
\label{heq}
\end{equation}
Introducing the dimensionless ratio $\alpha=\Delta/(2k_{\mathrm{B}}T_{\mathrm{L}})$ and the dimensionless counting field $\lambda=\xi k_{\mathrm{B}}T_{\mathrm{L}}$ we can write Eq. (\ref{CGF}) as (putting $\tau=t_0$)
\begin{equation}
F(\lambda)=-\frac{t_0 k_{\mathrm{B}}T_{\mathrm{L}}}{e^2R}\frac{\alpha}{\sinh \alpha}\left[\cosh \alpha-\sqrt{\sinh^2\alpha+\left(\frac{\pi \sin (\lambda \alpha)}{\alpha \sinh (\pi \lambda)}\right)^2}\right].
\label{CGF2}
\end{equation}
Importantly, the analytical form of the CGF as well as the corresponding distribution of the energy transfer $P_{t_0}(\varepsilon)$ can be understood in the following way: (i) First, for a given measurement there occur a large number $N\gg 1$ of in-and-out tunnelling events. Since the tunnelling is a quantum process, $N$ will fluctuate from measurement to measurement. The CGF $G(\chi)$ for the probability distribution $P(N)$ of the number of in-and-out events is directly obtained by substituting in Eq. (\ref{ME}) the off-diagonal elements $\Gamma_{\mathrm{in/out}}(\xi)\rightarrow \Gamma_{\mathrm{in/out}}^{(0)}e^{i \chi/2}$. Following the same procedure as above this gives
\begin{equation}
G(\chi)=-\frac{t_0 k_{\mathrm{B}}T_{\mathrm{L}}}{e^2R}\frac{\alpha}{\sinh \alpha}\left[\cosh \alpha-\sqrt{\sinh^2\alpha+e^{i\chi}}\right].
\label{CGF3}
\end{equation}
The square root form of the CGF in Eq. (\ref{CGF2}) is thus a consequence of the fluctuations in number of tunnel events $N$. (ii) Second, in every in-and-out tunnelling event, an energy $E$ is first transferred into the dot and then an energy $E'$ is transferred out of the dot. The probability distribution of the energies $E,E'$ is given from the integrand of the tunnel rates in Eq. (\ref{rates}), as 
\begin{equation}
p(E)=\frac{f(E+\Delta)-f(E)}{\Delta},
\label{singendist}
\end{equation}
and, taking into account that out-tunnelling for $E>0$ removes energy from the dot, $p(E')=p(-E)$, as plotted in Fig. \ref{curr}. Since the transferred energy in each event is independent on the number of events $N$, $P_{t_0}(\varepsilon)$ in Eq. (\ref{probdist}) can be written as the probability distribution for the energy transferred during $N$ statistically independent in-and-out tunnelling events, averaged with respect to the probability distribution for $N$. For the CGF $F(\lambda)$, this implies that we can make use of the cumulant composition rule and write 
\begin{equation}
F(\lambda)=G\left(\chi(\lambda)\right),
\label{CGFcomp}
\end{equation}
where the CGF $\chi(\lambda)$ for the energy transfer during a single in-and-out event is given from the moment generating function
\begin{eqnarray}
e^{i\chi(\lambda)}&=&\left(\int dE e^{i\lambda E/k_{\mathrm{B}}T_{\mathrm{L}}}p(E)\right)\left(\int dE e^{-i\lambda E/k_{\mathrm{B}}T_{\mathrm{L}}}p(E)\right) \nonumber \\
&=&h(\lambda)h(-\lambda)=\left(\frac{\pi \sin (\lambda \alpha)}{\alpha \sinh (\pi \lambda)}\right)^2
\label{CGFcomp2}
\end{eqnarray}
the product of the moment generating functions for energy transfer during in and out tunnelling. The function $h(\lambda)=h(\xi=\lambda/k_{\mathrm{B}}T_{\mathrm{L}})$ is given in Eq. (\ref{heq}). From Eqs. (\ref{CGF3}) - (\ref{CGFcomp2}) we see that we obtain the full CGF $F(\lambda)$ in Eq. (\ref{CGF2}). 

Having analysed the physical origin of $F(\lambda)$ in Eq. (\ref{CGF2}) we make some comment on its analytical structure. First, $F(\lambda)$ is an even function of $\lambda$, following from that $\sin(\lambda \alpha)/\sinh(\pi \lambda)$ is even in $ \lambda$. Thus, all odd cumulants of the total energy $\varepsilon$ are zero. As a consequence, the energy distribution $P_{t_0}(\varepsilon)$ is even in $\varepsilon$, around zero (since $T_\mathrm{e}=T_\mathrm{L}$). Second, $F(\lambda)$ is also an even function of $\alpha$, hence the sign of the charging energy difference $\Delta$ is not important for the energy transfer statistics. Third, in contrast to the statistics of number of events $N$ with the CGF $G(\chi)$ in Eq. (\ref{CGF3}), or the well studied charge transfer statistics for a system in the transport state \cite{Belzig,Esposito}, $F(\lambda)$ is not periodic in $\lambda$. This is a consequence of the fact that the energy transferred in each tunnelling event is not quantized.

\subsection{Cumulants and probability distribution}
Detailed information about the energy transfer is given by analysing the lowest order non-zero cumulants. Since all odd cumulants are zero, the average energy transfer $\langle \varepsilon \rangle=0$, i.e. zero average energy flow, as expected for thermal equilibrium. For the second cumulant we have 
\begin{equation}
\langle (\Delta \varepsilon)^2 \rangle=(k_{\mathrm{B}}T_{\mathrm{L}})^2\left.\frac{d^2F}{d(i\lambda)^2}\right|_{\lambda=0}=t_0\frac{2(k_{\mathrm{B}}T_{\mathrm{L}})^3}{3e^2R}\frac{\alpha(\pi^2+\alpha^2)}{\sinh (2\alpha)}.
\label{seccum}
\end{equation}
Making use of the composition relation in Eq. (\ref{CGFcomp}) and the fact that odd cumulants of $\chi(\lambda)$ are zero, we can also write
\begin{equation}
\langle (\Delta \varepsilon)^2 \rangle=(k_\mathrm{B}T_{\mathrm{L}})^2\left.\frac{dG}{d(i\chi)}\right|_{\chi=0} \left. \frac{d^2(i\chi)}{d(i\lambda)^2}\right|_{\lambda=0},
\end{equation}
with
\begin{equation}
\left.\frac{dG}{d(i\chi)}\right|_{\chi=0}=t_0\frac{k_{\mathrm{B}}T_{\mathrm{L}}\alpha}{e^2R\sinh(2\alpha)}=\frac{t_0}{\tau_\mathrm{E}},\hspace{0.5cm} \left. \frac{d^2(i\chi)}{d(i\lambda)^2}\right|_{\lambda=0}=\frac{2(\alpha^2+\pi^2)}{3}.
\label{rels}
\end{equation}
The energy fluctuations are thus equal to the fluctuations of the energy transferred during a single in-out tunnelling event, multiplied by the average number of events $\langle N\rangle=t_0/\tau_{\mathrm{E}}$ during the measurement. The number of events $\langle N \rangle$ is exponentially suppressed for increasing $\alpha=\Delta/2k_{\mathrm{B}}T_{\mathrm{L}}$, i.e. in the Coulomb blockade regime. In contrast, the single event energy fluctuations increase with increasing $\alpha$, a consequence of large $\Delta$ making the energy distribution $p(E)$ in Eq. (\ref{singendist}) broader.

\begin{figure}[h]
\begin{center}
\includegraphics[width=0.42\linewidth]{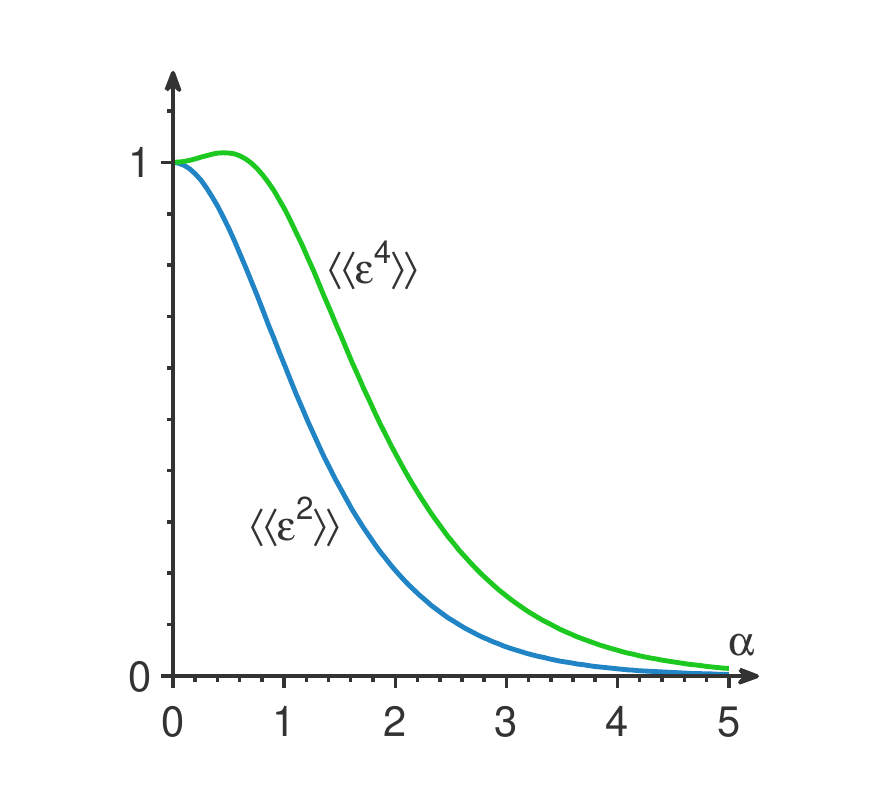}
\includegraphics[width=0.42\linewidth]{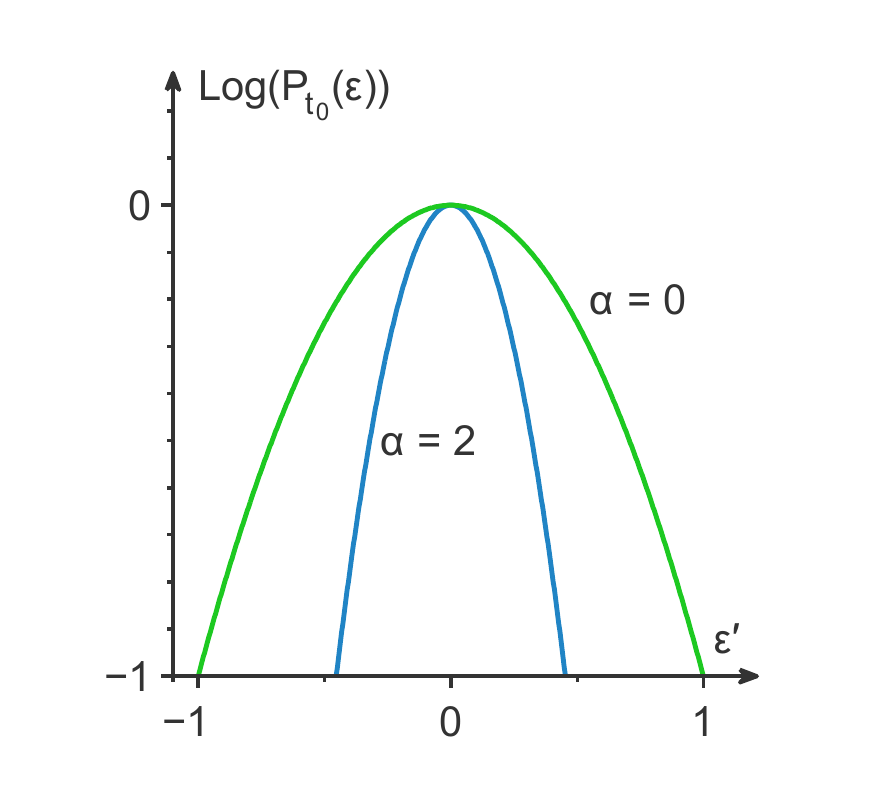}

\end{center}
\caption{Left: Second, $\langle \! \langle \varepsilon^2 \rangle \!\rangle$, and fourth,  $\langle \! \langle \varepsilon^4 \rangle \!\rangle$, cumulants as a function of $\alpha=\Delta/(2k_{\mathrm{B}}T_{\mathrm{L}})$, normalized to their respective values at $\alpha=0$. Right: Logarithm of probability distribution $P_{t_0}(\varepsilon)$ for two different $\alpha=0,2$, as a function of $\varepsilon ` = \varepsilon/\sqrt{2\langle \! \langle \varepsilon^2 \rangle \! \rangle|_{\alpha=0}}$.} 
\label{probdistfig}
\end{figure}

Higher-order cumulants provide information about the deviation from a Gaussian distribution. While the odd, third cumulant is zero, the fourth cumulant is nonzero and given by
\begin{equation}
\langle \! \langle \varepsilon^4 \rangle \!\rangle=\langle (\Delta \varepsilon)^2 \rangle \frac{(k_{\mathrm{B}}T_{\mathrm{L}})^2}{5\cosh^2\alpha}\left[-\alpha^2+\pi^2+(4\alpha^2+6\pi^2)\cosh(2\alpha)\right]
\end{equation}
We can also, similar to the second cumulant, write
\begin{equation}
\langle \! \langle \varepsilon^4 \rangle \!\rangle=(k_{\mathrm{B}}T_{\mathrm{L}})^4\left[3\left.\frac{d^2G}{d(i\chi)^2}\right|_{\chi=0}\left(\left.\frac{d^2(i\chi)}{d(i\lambda)^2}\right|_{\lambda=0}\right)^2+\left.\frac{dG}{d(i\chi)}\right|_{\chi=0} \left.\frac{d^4(i\chi)}{d\lambda^4}\right|_{\lambda=0}\right]
\end{equation}
where, in addition to the results in Eq. (\ref{rels}), we have
\begin{equation}
\left.\frac{d^2G}{d(i\chi)^2}\right|_{\chi=0}=\frac{t_0}{2\tau_\mathrm{E}}\frac{\cosh(2\alpha)}{\cosh^2\alpha}, \hspace{0.5cm} \left.\frac{d^4(i\chi)}{d\lambda^4}\right|_{\lambda=0}=\frac{4(\pi^4-\alpha^4)}{15}
\end{equation}
Thus, for $\langle \! \langle \varepsilon^4 \rangle \!\rangle$ we see that in addition to the fourth cumulant of the single event energy fluctuations multiplied by the average number of events $\langle N\rangle$, there is an additional contribution from the fluctuation in number of events times the square of the single event energy fluctuations. Note that total $\langle \! \langle \varepsilon^4 \rangle \!\rangle$, plotted in Fig. \ref{probdistfig}, has the same sign, independent of $\alpha$.

Turning to the full distribution $P_{t_0}(\varepsilon)$, Eq. (\ref{probdist}), for small energies $\varepsilon$ we can (due to $N\gg 1$) perturbatively evaluate the full distribution in the saddle point approximation. To exponential accuracy we have
\begin{equation}
\ln P_{\mathrm{t_0}}(\varepsilon)=-i\xi^*\varepsilon+F(\xi^*), \hspace{0.5cm} -i\varepsilon+dF/d\xi|_{\xi=\xi^*}=0
\end{equation}
which gives, to fourth order in energy 
\begin{equation}
\ln P_{t_0}=-\frac{1}{2\langle\!\langle \varepsilon^2 \rangle \! \rangle}\varepsilon^2+\frac{\langle\!\langle \varepsilon^4 \rangle \! \rangle}{24 \langle\!\langle \varepsilon^2 \rangle \! \rangle^4}\varepsilon^4
\label{cumexpPE}
\end{equation}
The logarithm of the distribution for two different $\alpha$ is plotted in Fig. \ref{probdistfig}. We note that for $\alpha \sim 1$ and below, the fourth order term $\propto \varepsilon^4$ in Eq. (\ref{cumexpPE}) is negligibly small, i.e. the distribution is effectively Gaussian.
 
To summarize the section, we find that for equal temperatures in the dot and the lead, $T_{\mathrm{e}}(t)=T_\mathrm{L}$, the energy transferred fluctuates from measurement to measurement, with only the average over many measurements being zero. The fluctuations are a consequence of both the fluctuations in energy transferred during a single in-and-out tunnelling event and the fluctuations in the number of events during the measurement.  

\section{Temperature fluctuations}

As a result of the total energy fluctuations, it is clear that in the quasi-equilibrium regime described by Eq. (\ref{timescales}) the electron temperature $T_{\mathrm{e}}(t)$ will develop fluctuations in time. The mechanism underlying the fluctuations was discussed above, with a typical time trace of the fluctuating temperature shown in Fig. \ref{fig_system}. Our aim is to derive a complete description of the statistics of the temperature fluctuations. Formally, considering a measurement time $t_{\mathrm{0}}$, the full distribution of the time-dependent temperature in the interval $[0,t_0]$ can be written \cite{ZJ} as a functional Fourier transform
\begin{equation}
P[T_{\mathrm{e}}(t)]=\int \mathcal{D}\zeta(t)\exp\left[-i\int_{0}^{t_{\mathrm{0}}} dt' T_{\mathrm{e}}(t')\zeta(t')+S[\zeta(t)]\right]
\label{funcformP}
\end{equation}
where the measure $\mathcal{D}\zeta(t)$ denotes integration over all possible time-dependent counting fields $\zeta(t)$ and $S[\zeta(t)]$ is the generating functional. Performing $n$ functional derivatives of $S[\zeta(t)]$ with respect to $\zeta(t)$ we get the $n$-point (irreducible) temperature correlation function in the time domain, as
\begin{equation}
\langle \! \langle  T_{\mathrm{e}}(t_1)... T_{\mathrm{e}}(t_n) \rangle \! \rangle = \frac{1}{i^n}\left.\frac{\delta^n S[\zeta(t)]}{\delta \zeta(t_1)...\delta \zeta(t_n)}\right|_{\zeta(t)=0}
\label{shorttcorr}
\end{equation}

Importantly, for a typical time $\tau_\mathrm{E}$ satisfying Eq. (\ref{timescales}), in the range $1$~ns to $1~\mu$s, resolving individual tunnelling events in time with existing experimental techniques is extremely challenging. Thus, the short time temperature correlators in Eq. (\ref{shorttcorr}) are difficult to investigate experimentally. Instead, the experimentally accessible quantities are the low-frequency correlators, in particular the magnitude of the fluctuations, or the temperature noise, given by
\begin{equation}
\langle (\Delta T_{\mathrm{e}})^2 \rangle = \frac{1}{t_0} \int_0^{t_0}\!\!\int_0^{t_0}dtdt'\langle \Delta T_{\mathrm{e}}(t)\Delta T_{\mathrm{e}}(t')\rangle,  \hspace{0.5cm} \Delta T_{\mathrm{e}}(t)=T_{\mathrm{e}}-\langle T_{\mathrm{e}}\rangle
\label{Tnoise}
\end{equation}
We note that the temperature noise can be written as $\langle (\Delta T_{\mathrm{e}})^2 \rangle=\langle \! \langle \theta^2 \rangle \! \rangle/t_0$ where have introduced the time-integrated temperature
\begin{equation}
\theta=\int_0^{t_0}T_{\mathrm{e}}(t)dt.
\label{Tdef}
\end{equation} 
This motivates us to focus our discussion and present analytical results for the fluctuations of $\theta$.  The full distribution of $\theta$, $P(\theta)$, can be written, similarly to Eq. (\ref{probdist}), in terms of a CGF $S(\zeta)$ as ($\zeta$ is constant in time)
\begin{equation}
P(\theta)=\frac{1}{2\pi}\int d\zeta e^{-i\zeta \theta +S(\zeta)}.
\label{Ptheta}
\end{equation} 
From $S(\zeta)$ the different cumulants are obtained by successive differentiation, in particular the average temperature $\langle T_{\mathrm{e}} \rangle=\langle \theta \rangle/t_0=(1/t_0)dS(\zeta)/d(i\zeta)|_{\zeta=0}$ and the fluctuations $\langle (\Delta T_{\mathrm{e}})^2 \rangle=\langle \! \langle \theta^2 \rangle\!\rangle/t_0=(1/t_0)d^2S(\zeta)/d(i\zeta)^2|_{\zeta=0}$. 

To provide a physically compelling picture of the temperature fluctuations, below we first discuss the average and the time-dependent second-order correlation function within a Boltzmann-Langevin approach \cite{Buttrev,Heikkilabook}. Then the generating functional $S[\zeta(t)]$, and hence the full distribution $P[T_{\mathrm{e}}(t)]$, Eq. (\ref{funcformP}), is derived within the framework of the stochastic path integral formalism \cite{pilgrambutt,SPI2}. We note that the stochastic path integral approach can be understood as a formal extension of the Boltzmann-Langevin approach to higher order correlators \cite{SPI2}. Moreover, it is equivalent to a fully quantum mechanical approach \cite{kinpil} taken in the semiclassical limit.  

\subsection{Boltzmann-Langevin approach}
As a starting point, we state the relation between the time-dependent temperature $T_{\mathrm{e}}(t)$ and the total energy of the dot $E(t)$, given by
\begin{equation}
E(t)=\nu_{\mathrm{F}} \int d\epsilon \epsilon \left[f(\epsilon,T_{\mathrm{e}}(t))-f(\epsilon,0)\right]=\nu_{\mathrm{F}} \frac{\pi^2k_{\mathrm{B}}^2T_{\mathrm{e}}^2(t)}{6}=\frac{C_{\mathrm{e}}(T_{\mathrm{e}}) T_{\mathrm{e}}(t)}{2}
\label{ETeq}
\end{equation}
where $\nu_{\mathrm{F}}$ the dot density of states at Fermi energy and we introduced $C_{\mathrm{e}}(T_{\mathrm{e}})=\nu_{\mathrm{F}}\pi^2k_{\mathrm{B}}^2T_{\mathrm{e}}/3$, the heat capacity of the free electron gas at temperature $T_{\mathrm{e}}$. We also note that the rate of change of the dot energy is, by definition, equal to the energy current flowing into the dot 
\begin{equation}
\frac{dE(t)}{dt}=j_{\mathrm{\varepsilon}}(t)
\label{encurrdef}
\end{equation}
At steady state, $dE(t)/dt=0$, we have $j_{\mathrm{\varepsilon}}(t)=j_{\mathrm{\varepsilon}}=0$. The expression for the steady state energy current $j_{\mathrm{\varepsilon}}=\langle \varepsilon \rangle/t_{\mathrm{0}}$ was derived below Eq. (\ref{gamman}). The condition of no energy current, $j_{\mathrm{\varepsilon}}=0$, is fulfilled for equal dot and reservoir temperatures, $T_{\mathrm{e}}=T_{\mathrm{L}}$.

Turning to the fluctuations, following the Langevin scheme, we write the energy current and the dot temperature as sums of a constant, time averaged part and a fluctuating part, as
\begin{equation}
j_{\mathrm{\varepsilon}}(t)=j_{\mathrm{\varepsilon}}+\Delta j_{\mathrm{\varepsilon}}(t)=\Delta j_{\mathrm{\varepsilon}}(t), \hspace{0.5cm} T_{\mathrm{e}}(t)=T_{\mathrm{e}}+\Delta T _{\mathrm{e}}(t)=T_{\mathrm{L}}+\Delta T_{\mathrm{e}}(t)
\label{fluctdef}
\end{equation}
The energy current fluctuations $\Delta j_{\mathrm{\varepsilon}}(t)$ can be written as 
\begin{equation}
\Delta j_{\mathrm{\varepsilon}}(t)=\delta j_{\mathrm{\varepsilon}}(t)+\Delta T_{\mathrm{e}}(t) \left. \frac{dj_{\mathrm{\varepsilon}}}{dT_{\mathrm{e}}}\right|_{T_{\mathrm{e}}=T_{\mathrm{L}}}=\delta j_{\mathrm{\varepsilon}}(t)-\kappa\Delta T_{\mathrm{e}}(t)
\end{equation}
where $\delta j_{\mathrm{\varepsilon}}(t)$ are the "bare" fluctuations caused by the stochastic nature of the electron tunnelling at the dot-reservoir contact and $\Delta T_{\mathrm{e}}(t) \kappa$ is induced by the fluctuating dot temperature. Here we introduced $\kappa \equiv \kappa(T_{\mathrm{L}})$, with $\kappa(T_{\mathrm{e}})=-d j_{\mathrm{\varepsilon}}/dT_{\mathrm{e}}$, by definition the thermal conductance of the dot-reservoir contact at the reservoir temperature $T_{\mathrm{L}}$. As a next step we combine the expressions for $dE(t)/dt$ from Eqs. (\ref{ETeq}) and (\ref{encurrdef}) and then insert the expressions for $j_{\mathrm{\varepsilon}}(t)$ and $T_{\mathrm{e}}(t)$, from Eq. (\ref{fluctdef}), giving to leading order in the fluctuating quantities
\begin{equation}
C_{\mathrm{e}}\frac{d\Delta T_{\mathrm{e}}(t)}{dt}=
\delta j_{\mathrm{\varepsilon}}(t)-\kappa   \Delta T_{\mathrm{e}}(t) 
\label{dteq}
\end{equation}
where we write for short $C_{\mathrm{e}} \equiv C_{\mathrm{e}}(T_{\mathrm{L}})$, the heat capacity of the dot at the reservoir temperature $T_{\mathrm{L}}$. Fourier transforming Eq. (\ref{dteq}) and solving for $\Delta T_{\mathrm{e}}(\omega)$ we have
\begin{equation}
\Delta T_{\mathrm{e}}(\omega)=\frac{1}{i\omega C_{\mathrm{e}}+\kappa}\delta j_{\mathrm{\varepsilon}}(\omega)
\end{equation}
Importantly, at equilibrium the fluctuation-dissipation relation applied to heat transport \cite{AvPekola,Heikkilabook} states that, in the semiclassical limit $k_{\mathrm B}T_{\mathrm L} \gg \hbar \omega$,  
\begin{equation}
\langle \delta j_{\mathrm{\varepsilon}}(\omega) \delta j_{\mathrm{\varepsilon}}(\omega')\rangle=2k_{\mathrm{B}}T_{\mathrm{L}}^2\kappa\delta(\omega+\omega')
\end{equation}
assuming a white noise spectrum of the energy current fluctuations. We can then write the temperature correlator in frequency space
\begin{equation}
\langle \Delta T_{\mathrm{e}}(\omega) \Delta T_{\mathrm{e}}(\omega')\rangle=2k_{\mathrm{B}}T_{\mathrm{L}}^2\frac{\kappa}{C_{\mathrm{e}}^2\omega^2+\kappa^2}\delta(\omega+\omega')
\end{equation}
In the time domain this becomes 
\begin{equation}
\langle \Delta T_{\mathrm{e}}(t) \Delta T_{\mathrm{e}}(t+t')\rangle=\frac{k_{\mathrm{B}}T_{\mathrm{L}}^2}{C_{\mathrm{e}}}e^{-|t'|/\tau_{\mathrm{C}}}, 
\end{equation}
which is just Eq. (\ref{Ttcorr}). As pointed out above, this result is in the canonical form for fluctuations of a physical quantity, in quasi-equilibrium, around thermodynamical equilibrium \cite{Landau}.  

\subsection{Stochastic path integral formulation}

For an extension of the Boltzmann-Langevin approach to the full statistical distribution of temperature fluctuations, we turn to a stochastic path integral formalism. For consistency and completeness of our presentation, we give the derivation of the CGF in some detail. As a starting point we consider a time scale $\tau$ much longer than the tunnelling energy relaxation time $\tau_{\mathrm{E}}$ but short enough for the temperature to change only marginally, $\tau \ll \tau_{\mathrm{C}}$. First, starting at time $t=0$ with an energy $E_0$ in the dot, the  probability that the energy in the dot at time $\tau$ is $E_1$, i.e. that an energy $E_{10}=E_1-E_0$ has been transferred, is given by Eq. (\ref{probdist})
\begin{equation}
P(E_{10})=\frac{1}{2\pi}\int d\xi_{0} e^{-i\xi_0 (E_1-E_0)+\tau F(\xi_0,E_0)}
\end{equation}
where $\xi_0$ is the energy counting field. Extending the reasoning to $N$ time bins, we can write the conditional probability that the energy $E_{10}$ is transferred through the first time bin, $E_{21}$ during the second etc., as
\begin{eqnarray}
&&P(E_{10})P(E_{21})....P(E_{NN-1})=\frac{1}{(2\pi)^N}\int ... \int d\xi_{0}...d\xi_{N-1} \nonumber \\
&\times & \exp\left(\sum_{n=0}^{N-1}\left[-i\xi_{n}(E_{n+1}-E_{n})+\tau F(\xi_{n},E_{n})\right]\right)
\end{eqnarray}
The probability that at the end of the $N$ time bins, the total energy $E_{N}-E_0$ has been transferred is obtained by integrating over all intermediate energies $E_{n}$ with $n=1,...,N-1$, as
\begin{equation}
P(E_{N})=\int ... \int dE_1...d E_{N-1} P(E_{10})...P(E_{NN-1}). 
\end{equation}
The joint probability to have the electron temperatures $T_{n} = T_\mathrm{e}(n\tau)$ at each respective time $n\tau$, $n=0,...,N-1$, can now be written as the conditional probability
\begin{eqnarray}
P_{N}(T_0,...,T_{N-1})&=&\int .. \int dE_1..d E_{N-1} P(E_{10})...P(E_{NN-1}) \nonumber \\
&\times &\prod \limits_{n=0}^{N-1} \delta\left(T_{n}-\mathcal{T}(E_n)\right)
\end{eqnarray}
where the $\delta$-functions impose the relation between dot energy and temperature, Eq. (\ref{ETeq}), at each time and we introduced $\mathcal{T}(E)$ defined from the relation $E=C_{\mathrm{e}}(\mathcal{T})\mathcal{T}/2$. Note that the probability distribution $P(E_0)$ for the initial energy $E_0$ will not be of importance in the long-time limit we are interested in. Writing the $\delta$-functions as Fourier transforms over the variables $\zeta_0,...,\zeta_{N-1}$ we obtain
\begin{equation}
P_{N}(T_0,...,T_{N-1})=\frac{1}{(2\pi)^N}\int .. \int d\zeta_\mathrm{0} ..d\zeta_{N-1} e^{-i\sum \limits_{n=0}^{N-1}\tau \zeta_n T_n +S_{N}(\zeta_0,...,\zeta_{N-1})}
\label{PT1TN}
\end{equation}
with the CGF $S_{{N}}(\zeta_0, ... ,\zeta_{{N-1}})$ given by
\begin{eqnarray}
&& e^{S_{N}(\zeta_0, ... ,\zeta_{N-1})}=\frac{1}{(2\pi)^N}\int .. \int dE_0..d E_{N-1} d\xi_0...d\xi_{N-1} \nonumber \\
& \times & \exp\left(\sum_{n=0}^{N-1}\Big[-i\xi_{{n}}(E_{{n+1}}-E_{{n}})+\tau F(\xi_{{n}},E_{{n}})+i\tau \zeta_{n} \mathcal{T}(E_n)\Big]\right)
\label{multiCGF}
\end{eqnarray}
Using the fact that the total energy/temperature only changes a small amount in every time bin we can now take the continuum limit of Eq. (\ref{PT1TN}), giving the expression in Eq. (\ref{funcformP}) for the full probability distribution $P[T_{\mathrm{e}}(t)]$, with the generating functional $S[\zeta(t)]$ expressed in terms of a stochastic path integral as
\begin{eqnarray}
e^{S[\zeta(t)]}&=&\int \!\!\! \int \! \mathcal{D} E(t)\mathcal{D}\xi(t) \exp\left(H[\xi,E] \right)
\label{genfcnal}
\end{eqnarray}
where 
\begin{equation}
H[\xi,E] = \int_0 ^t dt' \left(-i\xi(t')\frac{dE(t')}{dt'}+F[\xi(t'),E(t')]+i\zeta(t') \mathcal{T}(t')\right)
\end{equation}
and $\mathcal{D} E(t) , \mathcal{D} \xi(t)$ denote integration over all paths in energy and counting field space, respectively. 

\subsection{Saddle point solution and generating functional expansion}
In the long measurement time limit considered, with a large number of tunneling events, we can solve the path integral in the saddle point approximation, giving
\begin{equation}
S[\zeta(t)] = H[\xi^*(t),E^*(t)] 
\label{SPA equation}
\end{equation}
where $\xi^*(t)$ and $E^*(t)$ are the solutions of the saddle point equations $\delta H[\xi,E]/\delta \xi=0$ and $\delta H[\xi,E]/\delta E=0$. The saddle point equations are conveniently written in terms of $\xi^*$ and $\mathcal{T}^*=\mathcal{T}(E^*)$ as
\begin{eqnarray}
2iC_{\mathrm{e}}(\mathcal{T}^*)\frac{d\mathcal{T}^*(t)}{dt}&=&\frac{\partial F(\xi^*(t),\mathcal{T}^*(t))}{\partial \xi^*(t)}  \\
 -2iC_{\mathrm{e}}(\mathcal{T}^*)\frac{d\xi^*(t)}{dt}&=&\frac{\partial F(\xi^*(t),\mathcal{T}^*(t))}{\partial \mathcal{T}^*(t)} + i\zeta(t)
\end{eqnarray}
where we note that $C_{\mathrm{e}}(\mathcal{T}^*)=\nu_{\mathrm{F}}\pi^2k_{\mathrm{B}}^2\mathcal{T}^*(t)/3$.

It is not possible to solve these saddle point equations analytically in the general case and hence, we do not have a simple expression for the generating functional $S[\zeta(t)]$. However, in order to find the temperature correlation functions in Eq. (\ref{shorttcorr}) we can expand the generating functional order by order in $\zeta(t)$. This is done by solving the saddle point equations for $\xi^*(t)$ and $\mathcal{T}^*(t)$ order by order in $\zeta(t)$, inserting the solutions into $S[\zeta(t)]$ and then performing a functional expansion of $S[\zeta(t)]$ with respect to $\zeta(t)$. Here we do not provide any details of this calculation, note however that we expand $\xi^*(t)=\xi_0(t) +\int \limits_{-\infty}^\infty d\tau \xi_1(t-\tau) \zeta(\tau)+...$ and similarly for $\mathcal{T}^*(t)$, formally a functional expansion with time-dependent coefficients $\xi_n(t),\mathcal{T}_n(t)$. For the first two correlation functions we find
\begin{equation}
\langle T_{\mathrm{e}}(t) \rangle=-i
\frac{\delta S[\zeta]}{\delta \zeta(t)}\Bigg|_{\zeta =0} = T_\mathrm{L}, 
\label{C1}
\end{equation}
and
\begin{equation}
\langle \! \langle T_{\mathrm{e}}(t)T_{\mathrm{e}}(t') \rangle \! \rangle = -
 \frac{\delta^2 S[\zeta]}{\delta \zeta(t)\delta \zeta(t')}\Bigg|_{\zeta =0} = \frac{F_{\xi\xi}}{2 F_{\xi T}^2 \tau_{\mathrm C}}e^{-|t-t'|/\tau_{\mathrm{C}}}
\label{C2}
\end{equation}
where the correlation time $\tau_{\mathrm{C}} = -C_{\mathrm{e}}(T_{\mathrm{L}})/F_{\mathrm{\xi T}}$ and we have introduced
\begin{eqnarray}
F_{\mathrm{\xi\xi}}&=&\frac{1}{t_0}\left.\frac{d^2F}{d(i\xi)^2}\right|_{\xi=0,T_{\mathrm{e}}=T_\mathrm{L}}=\frac {\langle (\Delta \varepsilon)^2 \rangle}{t_0}, \nonumber \\
F_{\mathrm{\xi T}}&=&\frac{1}{t_0}\left.\frac{d^2F}{d(i\xi) dT}\right|_{\xi=0,T_{\mathrm{e}}=T_{\mathrm{L}}}=-\frac{\langle (\Delta \varepsilon)^2 \rangle}{2t_0k_{\mathrm{B}}T_{\mathrm L}^2}
\label{Fxt and Fxx expressions}
\end{eqnarray}
Here, for $F_{\mathrm{\xi\xi}}$ we used the result in Eq. (\ref{seccum}) and $F_{\mathrm{\xi T}}$ was evaluated directly from Eq. (\ref{CGF}), performing the integrals over the relevant generalized tunnel rates $\Gamma_{\mathrm{in/out}}(\xi)$ differentiated with respect to $\xi$ and $T$. Importantly, using the definition $\kappa=-dj_{\mathrm{\varepsilon}}/dT_{\mathrm{e}}|_{T_{\mathrm{e}}=T_{\mathrm{L}}}=-F_{\mathrm{\xi T}}$ we see that the results in Eqs. (\ref{C2}) and (\ref{Fxt and Fxx expressions}) are in agreement with the results obtained by the Boltzmann-Langevin approach and the fluctuation dissipation relation. In addition, the relation $F_{\mathrm{\xi T}}=-F_{\mathrm{\xi\xi}}/(2k_{\mathrm{B}}T_{\mathrm{L}}^2)$ together with Eq. (\ref{seccum}) provides us with an explicit expression for the thermal conductance at equilibrium,
\begin{equation}
\kappa=\frac{\langle (\Delta \varepsilon)^2 \rangle}{2t_0k_{\mathrm{B}}T_{\mathrm L}^2}=\frac{k_{\mathrm{B}}^2T_{\mathrm{L}}}{3e^2R}\frac{\alpha(\pi^2+\alpha^2)}{\sinh (2\alpha)}.
\end{equation}
We note that at $\alpha=0$, lifted Coulomb blockade, the thermal conductance is given by $\kappa={\mathcal L}_0T_{\mathrm L}/(2R)$, with  ${\mathcal L}_0=\pi^2k_{\mathrm B}^2/(3e^2)$ the Lorenz number. This can be understood a Wiedemann-Franz relation \cite{Zianni,Kubala} between the heat conductance $\kappa$ and the effective series conductance $1/(2R)$ for the in-and-out tunnelling events (recall that no net charge transport takes place).  

Higher-order correlation functions can be obtained using the same procedure, however the expressions become long and difficult to analyse in a transparent way. We therefore instead focus on the low-frequency regime where simpler and also experimentally more relevant results can be obtained.

\subsubsection{Low-frequency statistics}
In the long-time, low-frequency limit, we can disregard the time-dependence of the energy $E(t)$ and the counting fields $\xi(t)$ and $\zeta(t)$ in Eq. (\ref{genfcnal}). Consequently, the functionals and path integrals turn into ordinary functions and integrals respectively and the probability distribution $P[T_{\mathrm{e}}(t)]$ can now be written as a distribution of the time-integrated temperature $\theta = \int \limits_0^{t_0} dt T_\mathrm{e}(t)$, as in Eq. (\ref{Ptheta}), with the CGF $S(\theta)$ given from
\begin{eqnarray}
e^{S(\theta)}=\int \!\!\! \int \! dE d\xi e^{H(\xi,E)}, \hspace{0.5cm} H(\xi,E)/t_0=i \zeta \mathcal{T}+F(\xi,E).
\end{eqnarray}
Evaluating the double integral in the saddle point approximation and, similar to the discussion above, expanding the resulting CGF order by order in $\zeta$, we arrive at two the lowest order cumulants, 
\begin{equation}
\langle \theta \rangle=t_0T_\mathrm{L}, \hspace{0.5cm} \langle \! \langle \theta^2 \rangle \! \rangle=\langle (\Delta \theta)^2 \rangle=t_0\frac{F_{\mathrm{\xi\xi}}}{F_{\mathrm{\xi T}}^2},
\label{twoTcum}
\end{equation}
as expected from integrating the time-dependent correlation functions in Eqs. (\ref{C1}) and (\ref{C2}) over time. Writing out the total result explicitly, using Eqs. (\ref{seccum}) and (\ref{Fxt and Fxx expressions}), we have the low-frequency temperature fluctuations 
\begin{equation}
\frac{\langle \! \langle \theta^2 \rangle \! \rangle}{t_0}=\frac{6e^2RT_{\mathrm{L}}}{k_{\mathrm{B}}}\frac{\sinh(2\alpha)}{\alpha(\alpha^2+\pi^2)}=\frac{2k_{\mathrm{B}}T_{\mathrm{L}}^2}{\kappa}.
\label{Tfluc}
\end{equation}
We note that this function is monotonically increasing for increasing $\alpha$, i.e. stronger Coulomb blockade results in larger temperature fluctuations. However, since a large $\alpha \gg 1$ also leads to an exponentially large time $\tau_{\mathrm{E}}$ between tunnelling events, the requirement $\tau_{\mathrm{E}}\ll \tau_{\mathrm{e-ph}}$ in Eq. (\ref{timescales}) puts a physical upper boundary on the magnitude of the temperature fluctuations. 

For the third cumulant $\langle \! \langle \theta^3 \rangle \!\rangle$, the skewness of the probability distribution, we can proceed in a similar way as for the second one. The resulting cumulant can be written
\begin{equation}
\frac{\langle \! \langle \theta^3 \rangle \! \rangle}{t_0}=-\frac{3F_{\mathrm{\xi\xi}}}{F_{\mathrm{\xi T}}^5}\left[F_{\xi T T}F_{\xi\xi}-F_{\xi \xi T}F_{\xi T}\right]
\label{thirdcumexp}
\end{equation}
where $F_{\xi\xi},F_{\xi T}$ are given in Eq. (\ref{Fxt and Fxx expressions}) and
\begin{eqnarray}
F_{\mathrm{\xi\xi T}}&=&\frac{1}{t_0}\left.\frac{d^3F}{d(i\xi)^2 dT}\right|_{\xi=0,T_{\mathrm{e}}=T_\mathrm{L}}=\frac{1}{t_0}\frac{k_{\mathrm{B}}^3T_{\mathrm{L}}^2}{e^2R}  \frac{2\alpha\left[\alpha(\alpha^2+\pi^2)\coth(2\alpha)+\pi^2\right]}{3\sinh(2\alpha)}  , \nonumber \\
F_{\mathrm{\xi T T}}&=&\frac{1}{t_0}\left.\frac{d^3F}{d(i\xi) dT^2}\right|_{\xi=0,T_{\mathrm{e}}=T_{\mathrm{L}}}=\frac{-1}{t_0}\frac{k_{\mathrm{B}}^2}{e^2R} \frac{2\alpha ^2\left[(\alpha^2+\pi^2)\coth(2\alpha)-\alpha\right]}{3\sinh(2\alpha)} 
\label{Fxtt and Fxxt expressions}
\end{eqnarray}
Inserting the expressions for $F_{\xi\xi},F_{\xi T},F_{\xi\xi T}$ and $F_{\xi T T}$ into Eq. (\ref{thirdcumexp}) we arrive at
\begin{eqnarray}
\frac{\langle \! \langle \theta^3 \rangle \!\rangle}{t_0}&=&-\frac{e^4R^2T_{\mathrm{L}}}{k_{\mathrm{B}}^2}\frac{108\sinh(2\alpha)}{\alpha^2(\alpha^2+\pi^2)^3} \nonumber \\
&\times& \left[\alpha(\alpha^2+\pi^2)\cosh(2\alpha)-(2\alpha^2+\pi^2)\sinh(2\alpha)\right].
\end{eqnarray}

\begin{figure}[h]
\begin{center}
\includegraphics[width=0.8\linewidth]{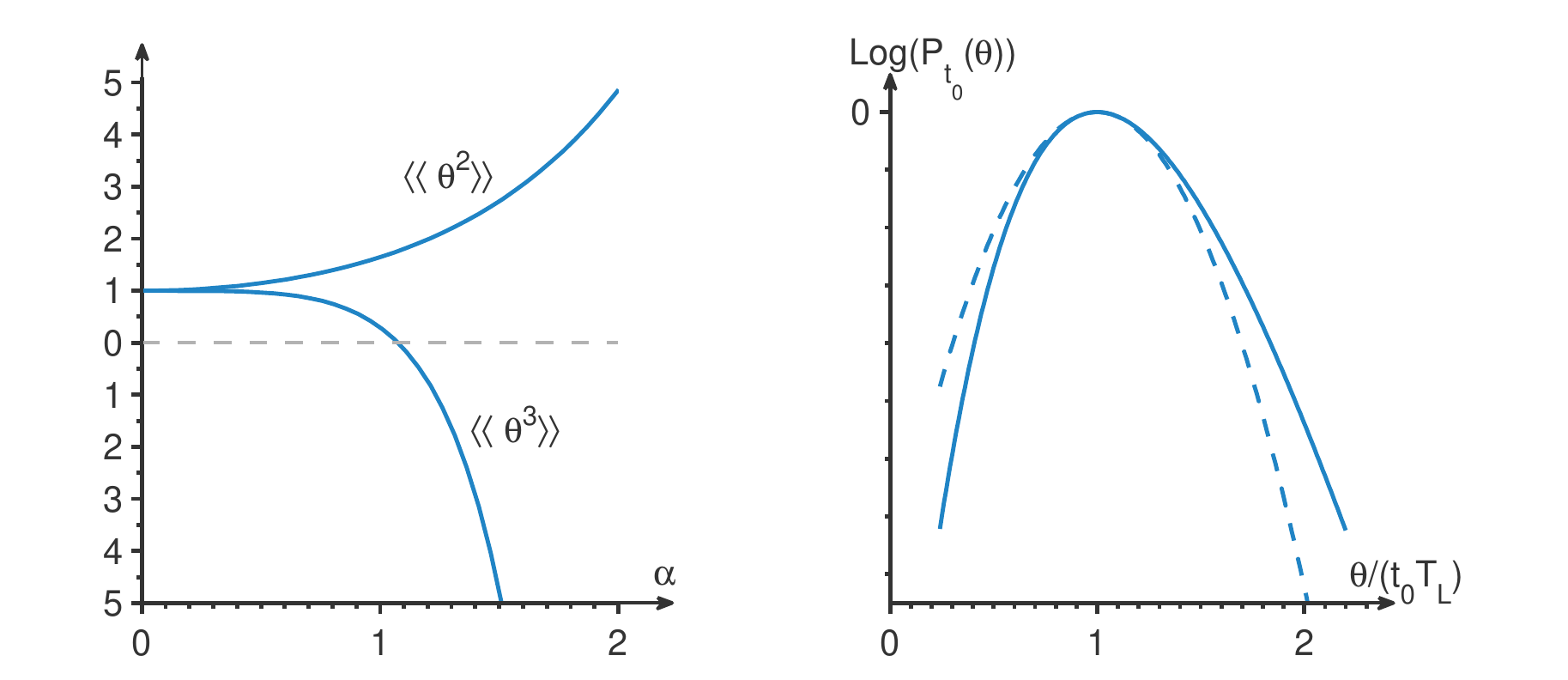}
\end{center}
\caption{Left: Second, $\langle \! \langle \theta(\alpha)^2 \rangle \! \rangle/\langle \! \langle \theta(0)^2 \rangle \! \rangle$, and third, $\langle \! \langle \theta(\alpha)^3 \rangle \! \rangle/\langle \! \langle \theta(0)^3 \rangle \! \rangle$, cumulant as a function of $\alpha=\Delta/2k_{\mathrm{B}}T_{\mathrm{L}}$, normalized with the value at $\alpha=0$. Right: Logarithm of probability distribution $P_{t_{0}}(\theta)$ (solid line) and Gaussian approximation (dashed line) for $\alpha=0$ (arb. units). The probability distribution $P_{t_{0}}(\theta)$ shows clear non-Gaussian features with a non-zero, positive skewness.}
\label{Tdist}
\end{figure}

The third cumulant is plotted as a function of $\alpha$ in Fig. \ref{Tdist}. At $\alpha=0$, i.e. lifted Coulomb blockade, $\langle \! \langle \theta^3 \rangle \!\rangle/t_0=216e^4R^2T_{\mathrm{L}}/(k_{\mathrm{B}}^2\pi^4)$, non-zero and positive. Increasing $\alpha$ the cumulant changes sign and increases rapidly in magnitude for increasing $\alpha$, i.e. for strong Coulomb blockade. 

To obtain a qualitative picture of the third cumulant we note that Eq. (\ref{thirdcumexp}) can be written
\begin{equation}
\frac{\langle \! \langle \theta^3 \rangle \! \rangle}{t_0}=
-\frac{3\langle (\Delta \varepsilon)^2\rangle}{\kappa^5}\left[\langle (\Delta \varepsilon)^2\rangle\frac{d\kappa}{dT_{\mathrm{e}}}-\kappa\frac{d\langle (\Delta \varepsilon)^2\rangle}{dT_{\mathrm{e}}} \right]
\label{thirdcumexpII}
\end{equation}
with the expression evaluated at $T_{\mathrm{e}}=T_{\mathrm{L}}$. That is, the fluctuations of $T_{\mathrm{e}}$ induce fluctuations of the heat conductance and the magnitude of the fluctuations of the transferred energy (on the time scale $\tau_{\mathrm{C}}$) as $d\kappa=(d\kappa/dT_{\mathrm{e}})\Delta T_{\mathrm{e}}$ and $d\langle (\Delta \varepsilon)^2\rangle=(d\langle (\Delta \varepsilon)^2\rangle/dT_{\mathrm{e}})\Delta T_{\mathrm{e}}$. These secondary, or cascaded \cite{Nagaeev1,Nagaeev2}, fluctuations, are correlated with the temperature fluctuations and responsible for the third cumulant.  

Instead of considering higher order cumulants individually we plot in Fig. \ref{Tdist}, as an illustrative example, the full probability distribution $P_{t_0}(\theta)$ in the limit $\alpha=0$. The probability distribution is evaluated numerically, to exponential accuracy in the saddle point approximation. From the plot the non-Gaussian nature of the fluctuations is clearly seen: while the distribution eventually goes to zero for $\theta=0$ (i.e. minus infinity for the logarithm of $P_{t_0}(\theta)$), it has a long tail for large $\theta$ which becomes increasingly fat for increasing $\alpha$ (not shown), consistent with the large second and large, negative third cumulant for large $\alpha$. 

\section{Summary and experimental realization}

To summarize, we have presented a theoretical analysis of the energy and temperature fluctuations of a single electron box, a metallic dot in the Coulomb blockade regime tunnel coupled to an electronic reservoir. The focus has been on the quasi-equilibrium regime, where the dynamics of the dot electron distribution, and hence the temperature, is driven by single electron tunnelling. We have investigated the different mechanism for the fluctuations of the energy transfer and discussed their effect on the temperature fluctuations. 

As a key result, we have presented a prediction for the magnitude of the low-frequency temperature fluctuations, in Eq. (\ref{Tfluc}). To provide an estimate of the magnitude of the fluctuations we consider the case with lifted Coulomb blockade, $\alpha=0$, a temperature $T_{\mathrm{L}}=100~$mK and take a typical resistance $R=100~$k$\Omega$. Inserted into the expression for the fluctuations we have $\langle \! \langle \theta^2 \rangle \! \rangle \sim 10^8~{\mathrm{K^2s}}$, or equivalently, $\sqrt{\langle \! \langle \theta^2 \rangle \! \rangle} \sim 100~\mu$K$/\sqrt{\mathrm{Hz}}$. This is of the same order as the measurement sensitivity reported in the most recent experiment \cite{Gasparinetti,Viisanen}, suggesting that the predicted temperature fluctuations could be observed with existing techniques. 

We also mention that very recent, unpublished, data \cite{Comm} show a sensitivity down towards $\sim 10~\mu$K$/\sqrt{\mathrm{Hz}}$. Interestingly, in Ref. \cite{Gasparinetti}  it is claimed that a resolution of this magnitude would enable single shot detection of single microwave photons (at $20$~GHz). Consequently, for the single electron box at large Coulomb blockade, $\Delta/k_{\mathrm{B}}\sim 1$ K, where the in-and-out tunnelling process is slow, typically $\tau_{\mathrm{E}} \gg \tau_{\mathrm{e-ph}}$, individual electron tunnelling events at large energies $\sim \Delta$ could be detected, in single shot, by monitoring the dot electron temperature in real time. We conclude by noting that the measurement of the dot electron temperature (not discussed in our work) could be carried out in the same way as in existing experiments (see e.g. \cite{Koski}), by non-invasively tunnel couple the dot to a superconducting lead. 
 
\section{Acknowledgements}
We gratefully acknowledge J. Pekola and S. Gasparinetti for a critical reading and helpful comments on earlier versions of the manuscript. We also thank J. Pekola for sharing unpublished experimental results and theoretical analysis on a closely related problem. We acknowledge support from the Swedish National Science Foundation and the Lund Nanometer Structure Consortium.

\end{document}